\documentclass{aa}  
\usepackage{graphicx}
\usepackage{txfonts}
\usepackage[colorlinks=true,citecolor=blue]{hyperref}

\usepackage{xcolor}
\usepackage{xspace}
\usepackage{amsmath}
\usepackage{tabularx}
\usepackage{booktabs}
\usepackage{scalerel}
\usepackage{longtable}
\usepackage{threeparttable}
\usepackage{lineno}
\usepackage{placeins}

\newcommand{\orcidicon}[1]{}
\usepackage[normalem]{ulem}

\newcommand{\tess}{\textsc{TESS}\xspace}
\newcommand{\rev}[1]{{#1}}

\begin{document} 

\title{WD~1054-226 revisited: a stable transiting debris system}

\author{J.~Korth\inst{\ref{ugen},\ref{lund}}\orcidicon{0000-0002-0076-6239}
        \and     A.~J.~Mustill\inst{\ref{lund}}\orcidicon{0000-0002-2086-3642}
        \and
        H.~Parviainen\inst{\ref{ull}, \ref{iac}}\orcidicon{0000-0001-5519-1391}
        \and
        E.~Villaver\inst{\ref{iac}, \ref{ull}}
        \and 
        J.~W. Kuehne\inst{\ref{mcd}}
        \and
        V.~ J.~S.~B\'ejar\inst{\ref{iac}, \ref{ull}}
        \and
        Y.~Hayashi\inst{\ref{komaba}} 
        \and 
        N.~Abreu García\inst{\ref{iac}, \ref{ull}} 
        \and
        T.~Kagetani\inst{\ref{komaba}}
        \and
        K.~Kawauchi\inst{\ref{ritsu}}
        \and 
        L.~Livingston\inst{\ref{abc},\ref{naoj},\ref{sokendai}}
        \and
        M.~Mori\inst{\ref{abc},\ref{naoj}}
        \and
        G.~Morello\inst{\ref{IAA},\ref{palermo}}
        \and
        N.~Watanabe\inst{\ref{komaba}}
        \and
        I.~Fukuda\inst{\ref{komaba}} 
        \and
        K.~Ikuta\inst{\ref{hito}}
        \and
        I.~Bonilla-Mariana\inst{\ref{ull}}
        \and 
        E.~Esparza-Borges\inst{\ref{iac}, \ref{ull}}
        \and
        G.~Fern\'andez-Rodr\'iguez\inst{\ref{ull}, \ref{iac}}
        \and
        A.~Fukui\inst{\ref{Komaba2},\ref{iac}}
        \and
        S.~Geraldía-González\inst{\ref{ull},\ref{iac}}
        \and 
        J.~González-Rodríguez\inst{\ref{iac}, \ref{ull}}
        \and 
        K.~Isogai\inst{\ref{okayama},\ref{komaba}}
        \and
        N.~Narita\inst{\ref{Komaba2}, \ref{abc}, \ref{iac}}
        \and
        E.~Palle\inst{\ref{iac},\ref{ull}}
        \and
        A.~Peláez-Torres\inst{\ref{IAA}}
        \and 
        M.~Sánchez-Benavente\inst{\ref{iac},\ref{ull}}
      }

    \institute{
    Observatoire Astronomique de l’Universit\'{e} de Gen\`{e}ve, Chemin Pegasi 51b, 1290 Versoix, Switzerland\label{ugen}\\
    \email{judithkorth@gmail.com}
    \and
    Lund Observatory, Division of Astrophysics, Department of Physics, Lund University, Box 118, 22100 Lund, Sweden\label{lund}
    \and
    Departamento de Astrof\'isica, Universidad de La Laguna (ULL), E-38206 La Laguna, Tenerife, Spain \label{ull}
    \and
    Instituto de Astrof\'isica de Canarias (IAC), E-38205 La Laguna, Tenerife, Spain \label{iac}
    \and 
    McDonald Observatory, Fort Davis, TX 79734, USA\label{mcd}
    \and 
    Department of Multi-Disciplinary Sciences, Graduate School of Arts and Sciences, The University of Tokyo, 3-8-1 Komaba, Meguro, Tokyo 153-8902, Japan\label{komaba}
    \and
    Department of Physical Sciences, Ritsumeikan University, Kusatsu, Shiga 525-8577, Japan\label{ritsu}
    \and
    Astrobiology Center, 2-21-1 Osawa, Mitaka, Tokyo 181-8588, Japan\label{abc}
    \and
    National Astronomical Observatory of Japan, 2-21-1 Osawa, Mitaka, Tokyo 181-8588, Japan\label{naoj}
    \and
     Astronomical Science Program, Graduate University for Advanced Studies, SOKENDAI, 2-21-1, Osawa, Mitaka, Tokyo, 181-8588, Japan\label{sokendai}
     \and
    Instituto de Astrofísica de Andalucía (IAA-CSIC), Glorieta de la Astronomía s/n, Genil, E-18008 Granada, Spain\label{IAA}
    \and
    INAF-Palermo Astronomical Observatory, Piazza del Parlamento, 1, 90134 Palermo, Italy\label{palermo}
    \and
    Graduate School of Social Data Science, Hitotsubashi University, 2-1 Naka, Kunitachi, Tokyo 186-8601, Japan\label{hito}
    \and
    Komaba Institute for Science, The University of Tokyo, 3-8-1 Komaba, Meguro, Tokyo 153-8902, Japan\label{Komaba2}
    \and 
    Okayama Observatory, Kyoto University, 3037-5 Honjo, Kamogatacho, Asakuchi, Okayama 719-0232, Japan\label{okayama}
         }

\date{Received MMMM DD, YYYY; accepted MMMM DD, YYYY}
 
  \abstract
   {\rev{A growing number of white dwarfs (WDs) exhibit one or more signs of remnant planetary systems, including transits, infrared excesses, and atmospheric metal pollution.} WD~1054-226 stands out for its unique, highly structured, and persistent photometric variability.}
   {We aim to investigate the long-term stability and nature of the periodic signals observed in WD~1054-226 to better understand the origin and evolution of its transiting material.}
   {We analyse all available \tess light curves from Sectors~9, 36, 63, and~90 using Lomb–Scargle (LS), Box-Least-Squares (BLS), and Gaussian process (GP) periodogram analyses. We complement these with multiband, high-cadence ground-based photometry from LCOGT, MuSCAT2, ALFOSC, and ProEM to test for colour dependence and confirm the periodicities.}
   {We confirm the persistence of the \rev{previously-reported} 25.01~h and 23.1~min periodicities over a six-year baseline. The 25.01~h signal \rev{shows some temporal evolution}, while the 23.1~min dips \rev{are highly coherent on long timescales}. A transient 11.4~h feature, previously reported, is detected only in early \tess sectors and is absent in recent data. No significant colour dependence is found in the ground-based observations.}
   {The stability of both the 25.01~h and 23.1~min signals indicates a long-lived, dynamically sculpted debris structure around WD~1054-226. The lack of colour dependence implies high optical depth, consistent with an opaque, edge-on debris ring rather than an optically thin dust population. This makes WD~1054-226 a key laboratory for testing models of remnant planetary systems around white dwarfs.}
   

   \maketitle

\section{Introduction}

When a star similar to the Sun ends its life, it sheds its outer layers and contracts into a white dwarf (WD). 
While the inner planetary system is engulfed, outer planets and minor bodies can survive and later perturb one another, leading to material being scattered onto the WD \citep[e.g.,][]{2002ApJ...572..556D,2011MNRAS.414..930B,2012ApJ...761..121M,2014MNRAS.445.4175V,2018MNRAS.476.3939M,2021MNRAS.504.3375S,2024RvMG...90..141V}. 
Those that pass within the Roche limit are tidally disrupted, generating compact dusty and gaseous discs 
\citep{2003ApJ...584L..91J,2006Sci...314.1908G}. 
Heavy element ``pollution'' in WD atmospheres \rev{\citep{2003ApJ...596..477Z,2014A&A...566A..34K,2024ApJ...976..156O}}, along with infrared excesses \rev{consistent with circumstellar dust discs \citep{1987Natur.330..138Z,1990ApJ...357..216G,2005ApJ...635L.161R,2007ApJ...663.1285J,2019MNRAS.487..133W,2024A&A...688A.168M,2026arXiv260210070},} and \rev{circumstellar} gas emission \rev{\citep{2006Sci...314.1908G,2020MNRAS.493.2127M,2025RMxAA..61..154S}}, provide strong evidence for ongoing accretion of such material. 
Recent JWST/MIRI spectroscopy has revealed a wide variety of dust mineralogies, including tentative silica glass signatures, pointing to high-temperature processing and collisional replenishment of small grains \citep{2025ApJ...981L...5F}.

In rare cases, fragments can be observed directly transiting their WD host. 
The first and best-studied example, WD~1145+017, showed multiple dust clouds orbiting near the Roche limit with a period of $\approx4.5$~h \citep{2015Natur.526..546V,2024MNRAS.530..117A}. 
Since then, over a dozen similar systems have been reported \citep[e.g.,][]{2020ApJ...897..171V,2021ApJ...912..125G,2021ApJ...917...41V,2024MNRAS.530..117A,2025ApJ...980...56H,2025PASP..137g4202B,2025ApJ...992..167G}, demonstrating diverse activity levels and dynamical pathways. However, only \rev{six} of these systems have measured orbital periods. 
Theoretical models suggest that these phenomena \rev{(discs, transits and pollution)} arise from gravitational perturbations by surviving planets \citep{2011MNRAS.414..930B,2018MNRAS.476.3939M}, leading to either full tidal disruption of an asteroid or comet within the Roche limit or to alternative outcomes. 
For example, secular and scattering perturbations can trigger repeated partial disruptions of planets or asteroids \citep{2021MNRAS.508.5671L,2024ApJ...974..100K}, while \citet{2025MNRAS.537.2214L} showed that bodies scattered to orbital pericentres just outside the Roche limit may undergo tidal circularisation to short-period (\rev{$\approx$10~h--1~d}) orbits. They showed that this mechanism can account for the orbital periods observed in systems such as WD~1145+017.
A companion study suggested that tidal heating during this evolution could induce volcanism, producing dust-rich ejecta and distinctive occultations \citep{2025MNRAS.541..610L}.

WD~1054-226 is a nearby, metal-polluted WD that displays uniquely structured and persistent photometric variability \citep{2022MNRAS.511.1647F, 2024MNRAS.533.1756R}. 
As described in \cite{2022MNRAS.511.1647F}, WD~1054-226 shows quasi-continuous dimming events, repeating with remarkable precision on a 25.01~h period, but with evolving morphology and variability at multiple harmonics, particularly the 65th harmonic at 23.1~min. This variability, drifting transit features that suggest additional periodicities, and the absence of unocculted starlight have been interpreted by \citet{2022MNRAS.511.1647F} as produced by an opaque, edge-on debris structure occulting the WD, subject to perturbations from a nearby massive body, such as a large asteroid fragment. Recent studies also suggest that the 25.01~h signal may be consistent with a partially circularised planetesimal orbit \citep{2025MNRAS.537.2214L} or with volcanically active bodies shedding dust in a narrow ring \citep{2025MNRAS.541..610L}. 

The stability and regularity of WD~1054-226 distinguish it from other transiting debris systems, making it a rare laboratory for testing models of evolved planetary systems. The published observational baseline for this system extends only over four years. Such systems have been observed only recently, and little is yet known about their long-term evolution or stability\rev{, though the small number of systems so far known exhibit a range of temporal behaviours. The} transits of the archetypal system WD~1145+017 have disappeared\rev{, or at least reduced in depth to become undetectable,} after several years \citep{2024MNRAS.530..117A}. \rev{Like WD~1054-226, ZTF~J0328-1219 shows complex transit features with no obvious out-of-transit baseline, but its periodicities (around 9.9 and 11.2 hours) change both within and between \emph{TESS} sectors \citep{2021ApJ...917...41V}. SBSS~1232+562 has undergone several months-long dimming events, after the last of which a 14.8-hour periodicity briefly appeared \citep{2025ApJ...980...56H}. ZTF~J1944+4557 shows cleaner transits with a clear flat baseline between the transit events, although these are still more complex than the single transit per cycle (here 4.97 hours) seen for transits of planets; these transits were observed in 2023 and 2025 but temporarily disappeared in 2024 \citep{2025ApJ...992..167G}.} Long-term monitoring of more systems is therefore essential to constrain the typical timescales on which these systems change, if indeed they all do. 
Here we present new observations and analysis of WD~1054-226 aimed at constraining the nature of the transiting material and its role in the late stages of planetary system evolution.

\section{Observations}
\subsection{\tess photometry}
WD~1054-226 \rev{(TIC 415714190)} was observed by the Transiting Exoplanet Survey Satellite \citep[\tess;][]{2015JATIS...1a4003R} in Sector~9 at 2~min cadence, and Sectors~36, 63, and 90 at a 20-sec cadence. We used the publicly available Presearch Data Conditioning (PDC) light curves \citep{2012PASP..124.1000S,2012PASP..124..985S,2014PASP..126..100S} produced by the Science Processing Operations Center \citep[SPOC:][]{2016SPIE.9913E..3EJ} at NASA Ames Research Center, downloaded from the Mikulski Archive for Space Telescopes.\!\footnote{\url{https://mast.stsci.edu}.}. 

\subsection{Ground-based photometry}
\label{sec:observations.gb}

\begin{figure*}
    \centering    \includegraphics[width=\linewidth]{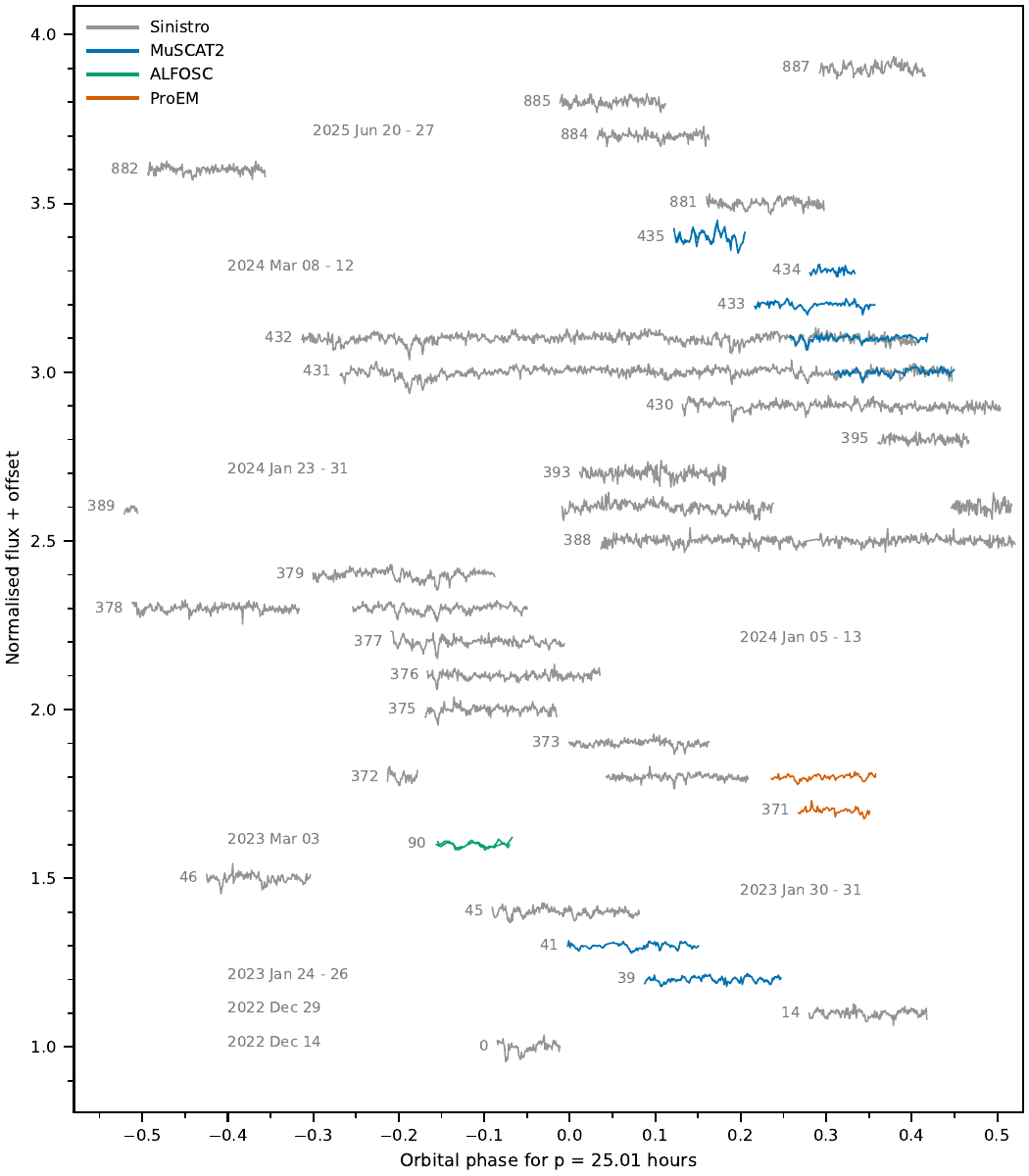}
    \caption{All $i$ band light curves ordered into groups, and phase-folded to the 25.01~h period, \rev{the reference time for phase folding is 2459928.85. The light curves of each group have a y-offset corresponding to the 25.01~h periodic signal cycle.}}
    \label{fig:lightcurves_i}
\end{figure*}

\begin{figure*}
    \centering
    \includegraphics[width=\linewidth]{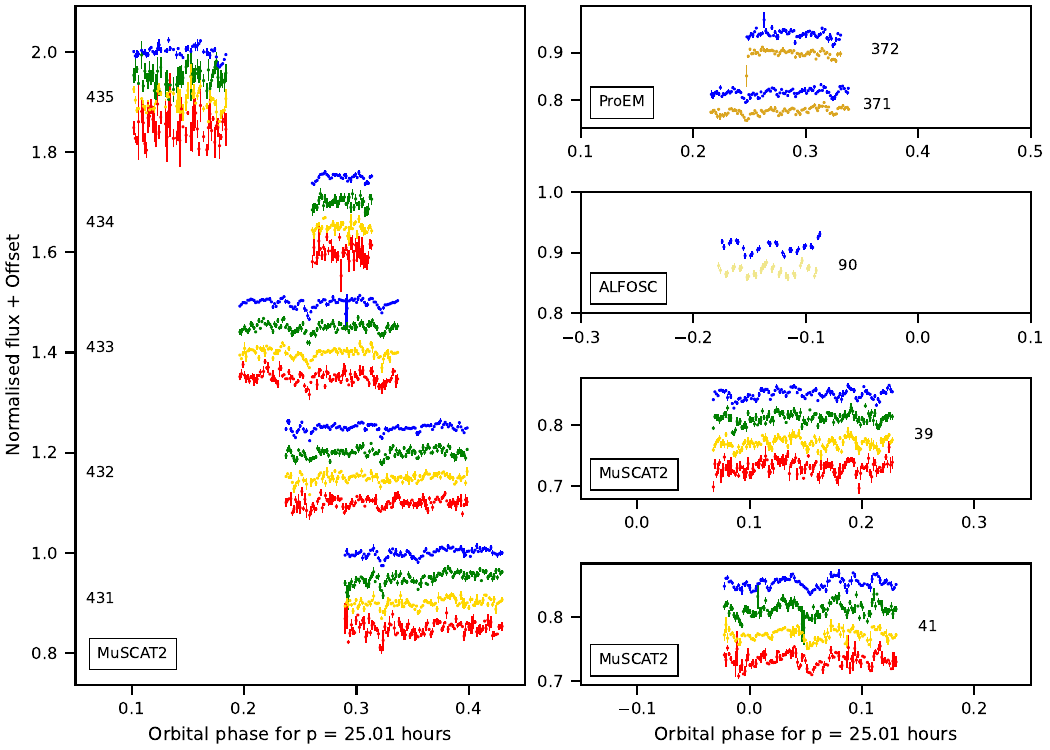}
    \caption{Multi-colour photometry from MuSCAT2, ProEM, and ALFOSC shown for filters $g'$ (blue), $r$ (green), $i$, and $i'$ (yellow), and $z_\mathrm{s}$ (red) and phase-folded to the 25.01-h period, \rev{the reference time for phase folding is 0.0. The MusCAT2 photometry is binned to 3-min, the ProEM photometry to 2-min, and the ALFOSC photometry is not binned. The light curves have a y-offset corresponding to the 25.01~h periodic signal cycle}.}
    \label{fig:lightcurves_multi}
\end{figure*}

\subsubsection{LCOGT/Sinistro}
We observed WD~1054-226 in the SDSS $i'$ band (617 to 894 nm) using the Sinistro cameras installed at 1-m telescopes from the Las Cumbres Observatory Global Telescope \citep[LCOGT;][]{2013PASP..125.1031B} between 2022-12-14 and 2025-06-27, covering observing windows from 2.13 to 7.48 hours. Details of the observations are reported in Table~\ref{tab:observation_summary} in Appendix~\ref{sec:observing_log}. The SDSS $i'$ band images were calibrated by the standard LCOGT \texttt{BANZAI} pipeline \citep{2018zndo...1257560M} and the photometry was reduced with our pipeline following standard photometry practices \citep{2019A&A...630A..89P}. The reduced $i'$ band photometry from LCOGT and all the other instruments is shown in Fig.~\ref{fig:lightcurves_i}.

\subsubsection{TCS/MuSCAT2}
We observed WD~1054-226 in $g'$ (400 to 550 nm), $r'$ (550 to 700 nm), $i$ (700 to 820 nm), and $z_s$ (820 to 920 nm) bands using the multi-band imager MuSCAT2 \citep{2019JATIS...5a5001N} mounted on the 1.5-m Telescopio Carlos Sánchez (TCS) at Teide Observatory, Spain between 2023-01-24 and 2024-03-12, with observing windows lasting from 2.0 to 3.7 hours. All the MuSCAT2 data were reduced by the MuSCAT2 pipeline \citep{2019A&A...630A..89P}. The pipeline performs standard calibrations (dark and flat field corrections) and aperture photometry. Three nights (2023-01-25, 2023-03-05, and 2023-03-31) were cloudy, and photometry for those nights is not included in the analysis. The multi-colour photometry from MuSCAT2, and all the other instruments, is shown in Fig.~\ref{fig:lightcurves_multi}, and the white light curve from MuSCAT2 with the colour differences is shown in Appendix~\ref{sec:m2_appendix} in Fig.~\ref{fig:M2_white}.

\subsubsection{Struve/ProEM}
We observed WD~1054-226 semi-simultaneously in SDSS $g'$ and $i'$ using the ProEM frame-transfer photometer mounted on the 2.1-m Otto Struve Telescope at McDonald Observatory, Texas, USA, on 2024-01-05 and 2024-01-06. The photometry was reduced with our pipeline in the same way as the LCOGT photometry.

\subsubsection{NOT/ALFOSC}
We observed WD~1054-226 semi-simultaneously in the SDSS $i'$ and $g'$ bands using the Alhambra Faint Object Spectrograph and Camera (ALFOSC) instrument installed at the 2.56-m Nordic Optical Telescope (NOT) at the Roque de los Muchachos Observatory on La Palma, Spain, on 2023-03-18. The photometry was reduced with our pipeline in the same way as the LCOGT photometry.

\section{Methods}
\subsection{Lomb-Scargle and Box-Least-Squares periodograms}

Previous searches for periodic signals in the photometry of WD~1054-226 have used Lomb-Scargle \citep[LS;][]{Lomb1976,Scargle1982} and Box Least Squares \citep[BLS;][]{Kovacs2002} periodograms \citep{2022MNRAS.511.1647F,2024MNRAS.533.1756R}. We repeat these analyses using both our ground-based photometry and the new \tess observations.

For the LS search, we use the \texttt{astropy} implementation to identify sinusoidal periodicities in both the \tess and ground-based data sets \rev{using the standard normalisation in the range 10~min to 50~days and 10~min to 3~days, respectively}. \rev{The False Alarm Probability (FAP) level is calculated using \texttt{astropy} implementation using the method described in \citet{Baluev2008}.} To search for periodic transit-like features in the \tess photometry, we employ \rev{the BLS algorithm implemented in} the Open Exoplanet Transit Search pipeline \citep[OpenTS;][]{2016MNRAS.461.3399P} using the \texttt{PyTransit} transit model \citep{Parviainen2015,2020MNRAS.499.1633P,2020MNRAS.499.3356P}. \rev{The BLS search was divided into short- and long-period searches, exploring trial periods of 0.007--0.2~days and 0.2--2~days, respectively.}

\subsection{Gaussian process \rev{period analyses}}
\label{sec:method.gp}
\subsubsection{Gaussian process kernel}

The LS periodogram is optimised for detecting sinusoidal variability, while the BLS periodogram is better suited for identifying periodic transit-like signals, where the duration of the dip is short compared to the period.

In this work, we adopt an additional approach and model the photometry using a Gaussian process \citep[GP;][]{Rasmussen2006} consisting of an aperiodic component and \rev{up to three quasi-periodic} components. \rev{The full covariance kernel is
\begin{equation}
\begin{split}
    k(\tau) &= \underbrace{\sigma_{ap}^2 \left(1 + \frac{\sqrt{3}\tau}{\rho}\right) \exp\left(-\frac{\sqrt{3}\tau}{\rho}\right)}_{\text{Aperiodic}} \\
    &\quad + \sum_{i=1}^{N} \underbrace{\sigma_{i}^2 \exp\left( -\frac{\tau^2}{2\ell_i^2} - \Gamma_i \sin^2\left( \frac{\pi \tau}{P_i} \right) \right)}_{\text{Quasi-periodic}}
\end{split}
    \label{eq:gp_kernel}
\end{equation}
where $\tau = |t - t'|$ is the time lag between data points and $N \in \{1, 2, 3\}$ is the number of periodic signals, and $\sigma$ is the amplitude of the component. The aperiodic variability is modelled with a Mat\'ern-3/2 covariance function, governed by the characteristic timescale $\rho$. The quasi-periodic signals are modelled by multiplying an exponential-sine-squared kernel with a squared-exponential kernel. This introduces an evolutionary timescale, $\ell$, which dictates the stability of the periodicity; a large $\ell$ ($\ell \gg P$) enforces a strictly periodic, persistent signal, while a smaller $\ell$ allows the signal's phase or amplitude to evolve or decay over time. The shape of the periodic feature is controlled by the characteristic inverse-length scale, $\Gamma$. Low values of $\Gamma$ constrain the periodic component to smooth, sinusoidal variations, while high values of $\Gamma$ allow for sharp, pulse-like features with rich harmonic content. This formulation allows us to model a wide range of behaviours, including long-timescale stochastic variability, sharp transit-like dips, and quasi-periodic signals with evolving morphology.}

\subsubsection{Period search with Gaussian processes}

\rev{We begin by carrying out period searches using the \tess data and the ground-based LCO data from January and March 2024. The searches are formulated as hypothesis tests between three competing models: $H_0$, in which the variability is explained entirely by an aperiodic process; $H_1$, in which the variability is explained by an aperiodic process and a single quasi-periodic signal with period $p_1$; and $H_2$, in which the variability is explained by an aperiodic process and two quasi-periodic signals with periods $p_1$ and $p_2$.}

\rev{We evaluate these hypotheses using Bayesian Information Criterion (BIC) periodograms. First, we fit the $H_0$ model by optimising the Mat\'ern-3/2 hyperparameters and computing the maximum log-likelihood. We then perform a period search by evaluating $H_1$ over a grid of trial periods $p_1$. At each trial period, we fix the aperiodic hyperparameters to their best-fit $H_0$ values and optimise the quasi-periodic kernel hyperparameters over a predefined grid. The resulting $\Delta\mathrm{BIC} = \mathrm{BIC}_{H_1} - \mathrm{BIC}_{H_0}$ identifies periods where the inclusion of a periodic component significantly improves the fit. \rev{A lower $\Delta\mathrm{BIC}$ indicates stronger evidence in favour of $H_1$, where $\Delta\mathrm{BIC} < -6$ can be considered strong evidence and  $\Delta\mathrm{BIC} < -10$ very strong \citep[][adopting $2\ln B_{10} \approx -\Delta \mathrm{BIC}$]{Kass1995}}. For $H_2$, we carry out a conditional search for secondary periodicities: we include a known primary period $p_1$ in the model and scan for an additional period $p_2$, allowing us to distinguish genuine secondary signals from harmonics of the dominant periodicity.}

\rev{We model the GPs using the \texttt{George} package \citep{Ambikasaran2016} with the basic solver rather than the HODLR approximation, so the computational cost is dominated by the $\mathcal{O}(N^3)$ scaling of covariance matrix inversion. To mitigate this, we adopt different strategies for the short- and long-period searches. For short periods (5--60~min), we do not bin the data in time but instead divide the light curves into chunks of at most $n$ data points and compute the total log-likelihood as the sum over the individual chunks. For long periods (3--29~h), we bin the \tess light curves into 22-minute intervals, keeping the total number of points per sector at around 1500; the ground-based light curves are binned into 4-minute intervals.}

\rev{The periodograms are calculated separately for each \tess Sector and separately for the LCO observations from January and March 2024. The searches are limited to at most two periodicities at each timescale, as our tests found no evidence for additional periodic signals.}

\subsubsection{Model refinement and comparison}

\rev{The BIC periodograms identify candidate periods but rely on grid searches with fixed aperiodic hyperparameters and approximate optimisation. To refine the results, we perform full GP optimisations for the most significant periods. For each candidate period or period pair, we optimise all hyperparameters jointly and recompute the BICs for $H_0$, $H_1$, and $H_2$. The optimisation uses the Differential Evolution \citep[DE;][]{Storn1997, Price2005} algorithm implemented in \texttt{PyTransit} \citep{Parviainen2015}. The resulting models are then compared to determine which set of periodicities best explains the data.}

\subsubsection{Posterior sampling}
\label{sec:method.gp.bics}

\rev{For the best-fitting models, we estimate the full posterior distributions of the GP hyperparameters using Markov Chain Monte Carlo (MCMC) sampling. We use the affine-invariant sampler implemented in \texttt{emcee} \citep{Foreman-Mackey2012}, initialising the walkers at the global posterior mode found by the DE optimisation. The sampling is carried out with 50 walkers for 5000 steps. We discard the first 4000 steps as burn-in and retain the final 1000 steps as the posterior sample.}

\subsection{Colour dependency}

Our ground-based photometry includes observations obtained in multiple passbands, either simultaneously or semi-simultaneously (see Sect.~\ref{sec:observations.gb}). We investigate whether the depths of the light-curve dips depend on colour, which could provide insights into the composition and size distribution of the material responsible for the dips \rev{\citep[e.g.,][]{2016A&A...589L...6A,2018MNRAS.474.4795X}}.  

The analysis parallels the approach of \citet{2022MNRAS.511.1647F}. First, we resample all passbands onto a common time grid to enable direct comparison; this step is required even for the MuSCAT2 data due to differences in exposure times. We then examine the correlations between fluxes across the passbands and assess whether they show significant non-linearity or deviations from unity, which would indicate a colour-dependent dip depth.

\section{Results}
\subsection{TESS  \rev{period search}}
We analysed all available \tess photometry using BLS, LS, and GP approaches.

\begin{figure*}
    \centering
    \includegraphics[width=\linewidth]{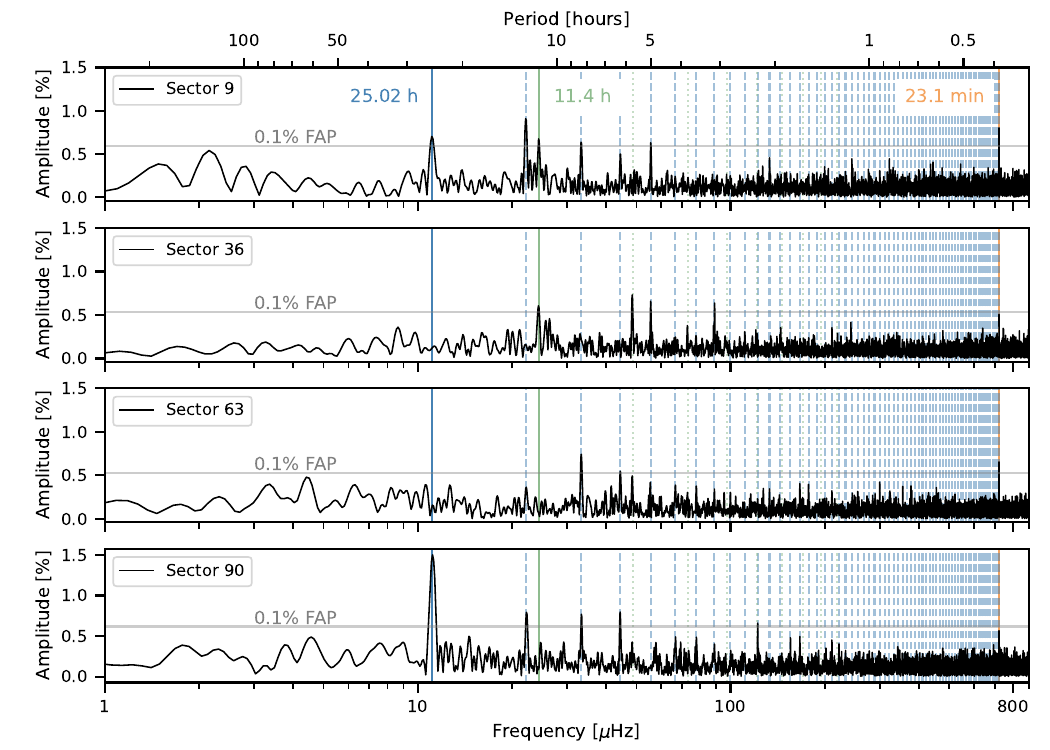}
    \caption{
    Lomb-Scargle periodograms of the \tess photometry from Sectors~9, 36, 63, and 90.  
    The known 23.1~min and 25.01~h periodicities are marked with solid \rev{orange and} blue lines, respectively, while the 11.4~h signal is marked with a solid green line.  
    Dashed \rev{and dotted} lines indicate harmonics of these signals using the same colour coding.}
    \label{fig:ls}
\end{figure*}

The BLS \rev{transit} search was performed per sector and across multiple sectors using both the 20~sec and 2~min cadence data. The \rev{long-period BLS search} identifies an 8.34~h signal in Sector 9, consistent with a harmonic of the 25.01~h modulation, as well as the 25.01~h periodicity itself in Sectors 36, 63, and 90, both individually and in the combined multi-sector searches. \rev{The short-period BLS search identifies the 23.1~min signal in all individual sectors, as well as in a search combining all the \tess Sectors.}

The LS search was performed using the 2~min cadence only; the periodograms (Fig.~\ref{fig:ls}) recover the two dominant periodicities previously reported by \citet{2022MNRAS.511.1647F}: the 23.1~min signal and the 25.01~h fundamental period.  
In addition, a secondary signal with a period of either 5.7~h or 11.4~h is detected in both Sectors~9 and~36, in agreement with the findings from \citet{2022MNRAS.511.1647F}.

\rev{The GP period searches were carried out separately for each \tess sector \rev{using the 2~min cadence \tess photometry}, computing  GP periodograms for long periods (3--29~h, using light curves binned into 22-minute intervals) and short periods (6--36~min, using unbinned photometry split into segments of 750 data points).}

\begin{figure*}
    \centering
    \includegraphics[width=\linewidth]{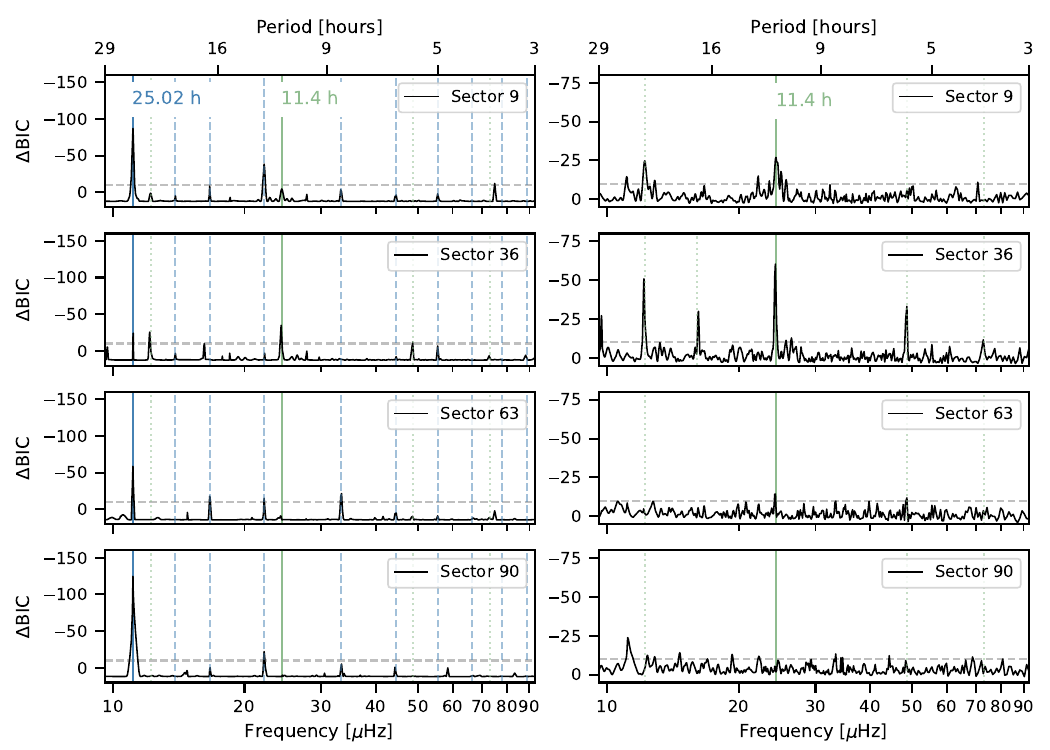}
    \caption{Gaussian process periodograms \rev{for the long-period searches (3--29~h) from the \tess Sector~9, 36, 63, and 90 photometry}. Left column: 1D periodograms for a single-period GP model.  
    The plotted signal amplitude $\Delta\mathrm{BIC}$ is defined as the difference between the BIC of the single-period GP model and that of the aperiodic GP model, i.e. $\Delta\mathrm{BIC}=\mathrm{BIC}_1(p_1)-\mathrm{BIC}_0$.  
    Right column: Conditional 1D periodograms for a two-period GP model where $p_1$ is fixed to 25.01~h.  
    Here, $\Delta\mathrm{BIC}=\mathrm{BIC}_2(p_1=25.01\mathrm{h},p_2)-\mathrm{BIC}_1(p_1=25.01\mathrm{h})$.  
    The known 25.01~h period and its harmonics are marked with solid and dashed blue lines, respectively, while the 11.4~h period and its harmonics are \rev{marked with the solid and dotted green lines}. \rev{The horizontal slashed line shows the $\Delta\mathrm{BIC}$ level of -10, which corresponds to very strong evidence in favour of the periodic signal \citep{Kass1995}.}}
    \label{fig:gps}
\end{figure*}

\rev{The long-period \tess results are shown in Fig.~\ref{fig:gps}. The GP periodograms recover the 25.01~h signal seen in the LS analysis, as well as a candidate periodicity at either 11.4 or 22.8~h in Sectors~9 and~36. The short-period search detects the known 23.1~min signal but finds no other significant features. To search for secondary periodicities, we computed conditional two-period GP periodograms (right column of Fig.~\ref{fig:gps}), comparing a one-period model with $p_1 = 25$~h against a two-period model with $p_1 = 25$~h and $p_2$ scanned across a grid of trial periods. In Sectors~9 and~36, the best-fitting models combine the 25.01~h signal with either the 11.4~h or 22.8~h periodicity: a weak secondary signal appears in Sector~9 and becomes stronger in Sector~36. By contrast, Sectors~63 and~90 are consistent with a single-period model dominated by the 25.01~h signal, with no significant secondary periodicities detected.}

Both the LS and GP analyses confirm that the known 25.01~h periodicity is the dominant, stable signal throughout the \tess observations, with little risk of it being an alias of another period.  However, the two methods produce slightly different results for the secondary signal in Sectors~9 and~36. The LS analysis identifies the strongest secondary peaks at 5.7~h and 11.4~h, whereas the GP analysis prefers periods of 11.4~h and 22.8~h.

\rev{To identify which long-period signals are genuinely present in each sector, we compared the BICs of competing GP models (Sect.~\ref{sec:method.gp.bics}): an aperiodic model, single-period models at 5.7, 11.4, 22.8, and 25.0~h, and two-period models combining the 25~h signal with each candidate secondary period. The resulting $\Delta\mathrm{BIC}$ values are summarised in Table~\ref{tab:bics}. The preferred model for Sectors~9 and~36 combines the 25.01~h and 11.4~h periodicities, while Sectors~63 and~90 are best explained by a single 25.01~h signal.}

\begin{table*}
    \centering
    \caption{Difference in BIC ($\Delta$BIC) relative to the best-fitting model for each \tess sector.}
    \begin{tabular*}{\textwidth}{@{\extracolsep{\fill}} r|r|rrrr|rrr}
    \toprule
    \toprule
     & & \multicolumn{4}{c|}{Single Period [h]} & \multicolumn{3}{c}{Two Periods [h] ($p_1 = 25$~h)} \\
     &  &  &  &  &  &  &  & \\
     Sector & Mat\'ern-3/2 & $p_1=5.7$ & $p_1=11.4$ & $p_1 = 22.8$ & $p_1 = 25.0$ & $p_2 = 5.7$ & $p_2=11.4$ & $p_2=22.8$ \\
    \midrule
    9 & 125.6 & 144.3 & 131.2 & 137.2 & 6.8 & 19.4 & \textbf{0.0} & 10.2 \\
    36 & 63.0 & 53.3 & 34.2 & 43.4 & 39.9 & 25.3 & \textbf{0.0} & 8.4 \\
    63 & 77.1 & 72.1 & 94.8 & 97.9 & \textbf{0.0} & 16.3 & 14.1 & 18.6 \\
    90 & 173.4 & 170.8 & 185.2 & 185.1 & \textbf{0.0} & 10.9 & 14.2 & 15.5 \\
    \bottomrule
    \end{tabular*}
    \tablefoot{A lower value of $\Delta$BIC indicates better agreement with the data.}
    \label{tab:bics}
\end{table*}

\subsection{Ground-based \rev{period search}}
 
When combining all available \rev{ground-based} datasets, the LS analysis detects only the 23.1~min modulation, while the longer-period signals remain undetected.  
This is likely due to the limited temporal coverage and the strongly non-sinusoidal nature of the variability, which makes the LS method less sensitive to the 25.01~h signal.  
In contrast, the GP periodogram \rev{combining the January~2024 and March~2024 observations}  recovers both the 23.1~min and 25.01~h modulations.

\begin{figure*}[t]
    \centering
    \includegraphics[width=1\linewidth]{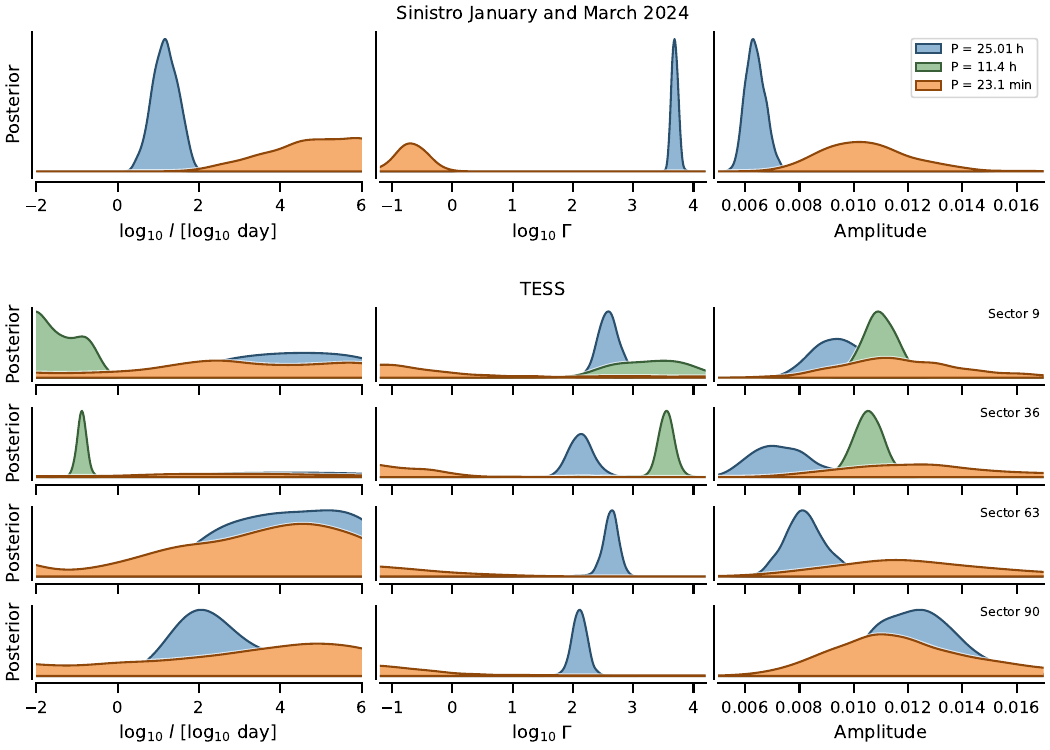}
    \caption{\rev{Posterior densities for the GP model hyperparameters estimated from the ground-based LCO observations and from separate \tess sectors. The first column shows the GP coherence scale, the second column shows the GP harmonic complexity, and the third column shows the amplitude of the periodicity. The 11.4~h period is estimated only from the \tess Sectors 9 and 36. The ground-based posterior estimates for the 25.01~h and 23.1~min periodicities are significantly better than the \tess ones due to the significantly higher photometric precision.}}
    \label{fig:gp_posteriors}
\end{figure*}

We also performed separate analyses for two data subsets obtained in January~2024 and March~2024.  
In January, the LS periodogram again identifies the 23.1~min modulation, with additional candidate long-period signals that are most likely spurious.  
\rev{The GP analysis confirms the presence of the 23.1~min variability and recovers the 25.01~h signal.}
A weaker feature near 6.25~h is also present, but this is consistent with being an alias of the 25.01~h modulation rather than an independent period.  
The March~2024 data show a similar picture: the 23.1~min signal is robustly detected in both LS and GP analyses. \rev{The GP analysis detects the 25.01~h modulation, but with a lower statistical significance than in the January~2024 data, and a tentative $\approx$5~h signal is also present but lacks statistical significance.}
Overall, the ground-based LS analysis consistently detects the 23.1~min modulation but fails to recover the 25.01~h periodicity, which is expected given its sensitivity to sinusoidal variability. 

\begin{table}[b!]
    \centering
    \caption{\rev{Period posterior estimates from the GP analyses.}}
    \begin{tabular*}{\linewidth}{@{\extracolsep{\fill}} lccc}
        \toprule
        Dataset & $p_1$ [h] & $p_2$ [h] & $p_3$ [min] \\
        \midrule
        \multicolumn{4}{l}{\textit{Ground-based}}\\[2pt]
        \quad J\&M  & $25.017 \pm 0.002$ & -- & $23.0947 \pm 0.0003$ \\
        \quad J  & $25.017 \pm 0.002$ & -- & $23.096 \pm 0.001$ \\
        \quad M  & $25.015 \pm 0.003$ & -- & $23.095 \pm 0.009$ \\[4pt]    
        \multicolumn{4}{l}{\textit{\tess sector}}\\[2pt]
        \quad 9    & $25.013 \pm 0.003$ & $11.47 \pm 0.11$       & $23.3 \pm 2.1$ \\
        \quad 36   & $24.997 \pm 0.007$ & $11.379 \pm 0.002$     & $22.9 \pm 2.0$ \\
        \quad 63   & $25.019 \pm 0.004$ & -- & $23.2 \pm 1.7$ \\
        \quad 90   & $25.012 \pm 0.004$ & -- & $22.7 \pm 1.9$ \\
        \bottomrule
    \end{tabular*}
    \tablefoot{The 11.4~h period is not constrained in \tess Sectors~63 and~90 or in the ground-based observations. For ground-based observations, J stands for an analysis using the January 2024 dataset, M for an analysis using the March 2024 dataset, and J\&M for a combined analysis with both datasets.}
    \label{tab:period_posterior}
\end{table}

\subsection{Gaussian process posterior analysis}

\begin{figure*}[h!]
    \centering
    \includegraphics[width=1\linewidth]{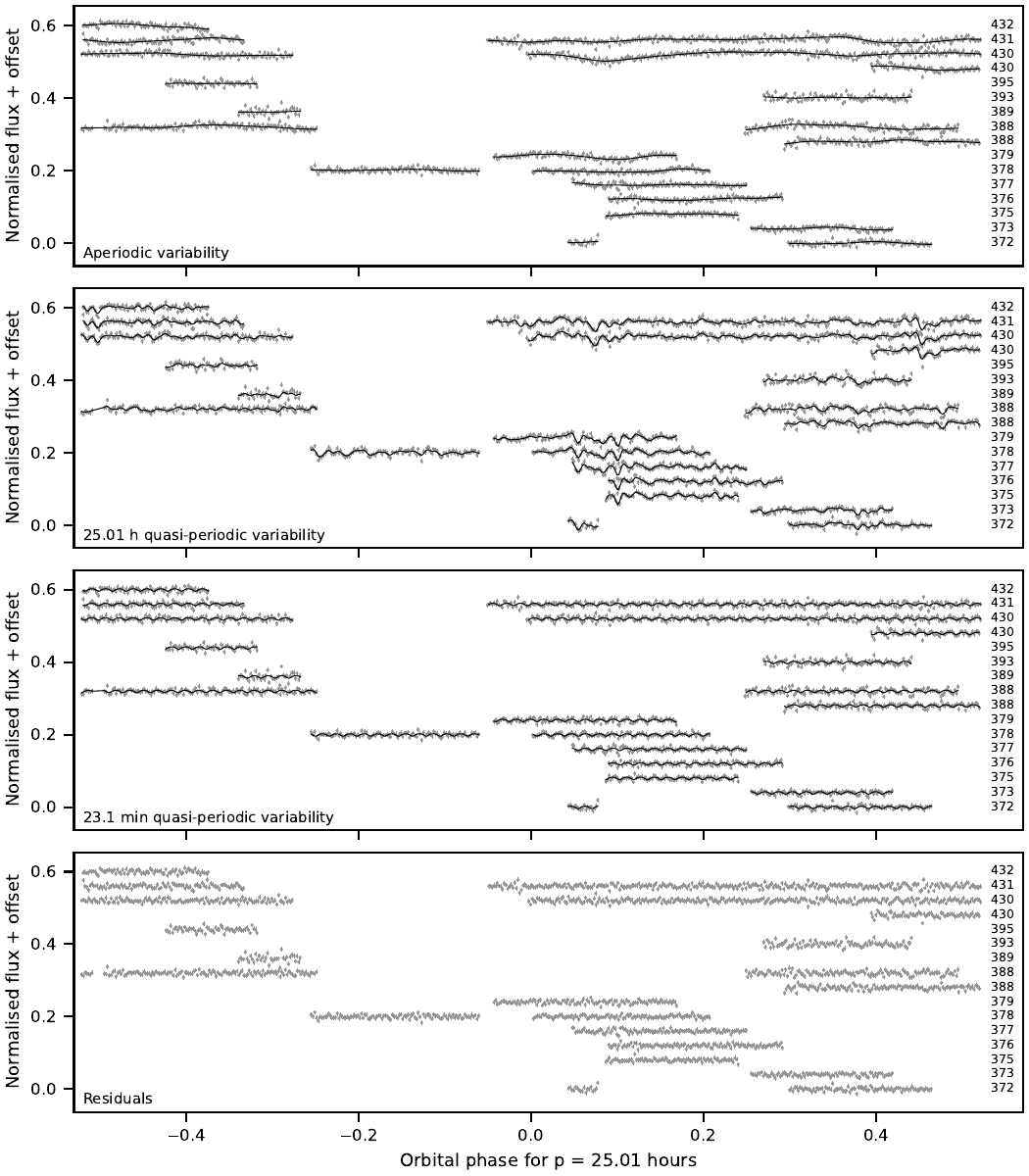}
    \caption{\rev{The ground-based LCO observations for January and March 2024 (points with error bars) together with the GP model consisting of an aperiodic component, a 25.01~h quasi-periodic component, and a 23.1~min quasi-periodic component (black line) folded over the 25.01~h period. The three panels from the top show each separate GP component with the other components removed from the observations and the model, and the bottom panel shows the observation residuals with respect to the full GP model. The cycle number is marked on the right, starting from the first light curve in the dataset.}}
    \label{fig:gp_model}
\end{figure*}

\rev{We present the GP period posteriors for the best models in Table~\ref{tab:period_posterior}, the remaining hyperparameter posteriors in Fig.~\ref{fig:gp_posteriors}, and the median posterior GP models for the ground-based observations in Fig.~\ref{fig:gp_model}. The period estimates for the 25.01~h and 23.1~min signals are consistent across all \tess sectors and the ground-based data. The two \tess sectors that exhibit the 11.4~h signal also agree within the uncertainties, but the signal is relatively weak in Sector~9. We do not find statistically significant drifter-like features in the ground-based data after removing the GP model, including the 25.01~h and 23.1~min signals.}

\rev{The hyperparameter posteriors (Fig.~\ref{fig:gp_posteriors}) reveal markedly different characteristics for each periodicity in the ground-based data. The 25.01~h signal has a well-constrained coherence timescale of $\ell \approx 10$~days (the 68\% central credible interval for $\log_{10} \ell$ is 0.85--1.5, which corresponds to an $\ell$ interval of 7--31~d) and very high harmonic complexity ($\log_{10}\Gamma \approx 4$), consistent with a sharp, complex periodic waveform that remains coherent within each observing campaign but evolves over the combined January--March baseline. The 23.1~min signal has low harmonic complexity ($\log_{10}\Gamma \approx -1$), indicating a smooth, nearly sinusoidal waveform, and a coherence timescale with a lower bound of $\ell > 170$~d (99\% credible lower limit), meaning it remains coherent over at least the full ground-based observing baseline. Its amplitude is comparable to that of the 25.01~h signal, at approximately 1\% of the normalised flux.}

\rev{In the \tess data, the 25.01~h signal shows poorly-constrained $\ell$ posteriors with $\ell \gtrsim 1$, indicating coherence over timescales likely longer than the sector baseline, with $\log_{10}\Gamma \approx 2$. The 11.4~h signal, detected only in Sectors~9 and~36, has a short coherence timescale ($\ell \lesssim 1$~day) and high harmonic complexity, as expected for transit-like dips with rapidly evolving morphology. The 23.1~min signal is poorly constrained by the \tess photometry, with broad posteriors for all hyperparameters, consistent with the low signal-to-noise ratio at this cadence for a faint target. Overall, the ground-based observations provide significantly tighter constraints on the hyperparameters of both the 25.01~h and 23.1~min signals than \tess, owing to their substantially higher photometric precision.}

\subsection{Colour dependence}

Finally, we investigated the wavelength dependence of the dips using the multi-colour photometry (Fig.~\ref{fig:lightcurves_multi}).  
We find no evidence of colour dependence in the dip depths within the observational uncertainties, consistent with previous studies of WD~1054-226 (Figs.~\ref{fig:M2_white} \rev{and \ref{fig:m2_zg}} in Appendix~\ref{sec:m2_appendix}). 

\section{Discussion}
As pointed out by \citet{2022MNRAS.511.1647F}, who analysed the \tess Sectors~9 and 36, large scatter prevents direct detection of transiting events but is beneficial for a frequency analysis. 
However, \citet{2024MNRAS.533.1756R} used a combined BLS and LS analysis and detected the 25.01~h period using Sectors~9, 36, and 63. We confirm this \rev{and the 23.1~min period, and find also that these signals persist in Sector~90}.

\begin{figure}
    \includegraphics[trim=0 0 0 25,clip, width=0.5\textwidth]{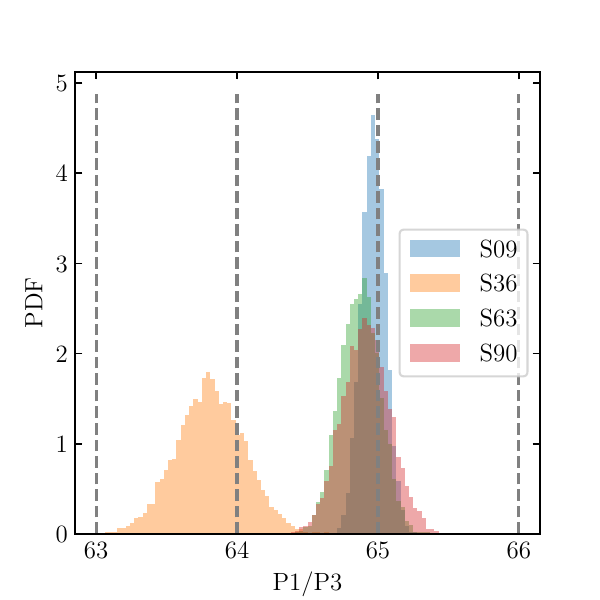}
    \caption{Posterior distribution of the ratio of the 25.01~h period and the 23.1~min period \rev{estimated from the 2~min cadence \tess photometry from the four \tess sectors}. All sectors lie close to the 65:1 commensurability except for Sector~36.}
    \label{fig:p_rat}
\end{figure}

The 25.01~h and 23.1~min signals are now largely persistent for over 6 years (2300 days, or $>$2200 cycles of the 25.01~h signal), indicating that they are reasonably long-lived and not rapidly decaying transients. This is consistent with a dynamical origin in which a disc is sculpted by a nearby massive perturber, such as an asteroid, whether via waves like in Saturn's rings or via resonant clumps. In this scenario, the 25.01~h period is that of the perturbing body, while the 23.1~min signal (at a 65:1 commensurability) is caused by resonant clumping of material, or waves launched by a 66:65 mean motion resonance at the disc edge, although both of these scenarios have qualitative challenges in explaining the shape of the light curves \citep[see discussion in][]{2022MNRAS.511.1647F}. Interestingly, the ratio of the 25.01~h and 23.1~min signals is very close to 65 in all \tess sectors except for Sector~36: here, the closest commensurability is 64:1 but the observed ratio lies further from the commensurability (Fig.~\ref{fig:p_rat}); note that the ``drifters'' were more noticeable in this sector than the others, and may affect the periodogram results. In these scenarios, the 25.01~h signal should remain \rev{phase-}coherent, as it is tied to the period and phase of the perturbing body. \rev{However, we determine a relatively low coherence time for this signal, which may arise from an evolution of its morphology rather than its phase.} If the disc is sufficiently massive and flat, relatively rapid migration of the perturbing body may take place on timescales much less than the expected disc lifetimes of $\sim1$\,Myr \citep{2023MNRAS.524....1V}. As an alternative scenario, the 25.01~h period may correspond to magnetically-trapped dust co-rotating with the stellar magnetosphere, as suggested for WD~1145+017 \citep{2017MNRAS.471L.145F} and for young main-sequence stars \citep{2024AJ....167...38B}. The stellar rotation period of $>3$ hours \citep{2022MNRAS.511.1647F} is consistent with this scenario. 

The exact nature of the 23.1~min signal remains \rev{undetermined}. In these dynamical scenarios, the 23.1~min signal should still show phase coherence, since the zero phase should be set by the longitude of conjunction with the perturbing body. The amplitude, however, could vary, for example, as the bodies composing the disc orbit at a different frequency from the perturbing body. 
\rev{The coherent 23-minute signal is consistent with a constant forcing frequency, for example, from an asteroid's mean motion resonances.}
A possible challenge to these dynamical models comes from the relatively wide orbit of the perturbing body. Its orbital period of 25.01~h is too large to have permitted tidal circularisation, and it may possess a considerable residual eccentricity in the range $0.2-0.4$ \citep{2025MNRAS.537.2214L}. This would likely render these dynamical models untenable, as such nearby mean motion resonances as the 66:65 will likely be destabilised by such a high eccentricity. However, other means exist of lowering a scattered body's eccentricity, such as interactions with any pre-existing disc: it is not necessary that the disc material actually originates from the perturbing body \citep{2018MNRAS.476.3939M}.

The lack of colour dependence of the transits is in line with observations of WD~1145+017 \rev{\citep{2016A&A...589L...6A,2018MNRAS.481..703I,2018MNRAS.474.4795X}, WD~J0328-1219 \citep{2024RNAAS...8..173G}, ZTF~J1944+4557 \citep{2025ApJ...992..167G}, and SBSS~1232+563 \citep{2025ApJ...980...56H},} and implies a lack of small grains in the structures transiting \rev{these stars}, if the structures are in an optically-thin regime \citep{2018MNRAS.474.4795X}. However, small grains should be present as they are expected to be rapidly replenished by collisions in these discs. \cite{2018MNRAS.474.4795X} argued for WD~1145+017 that small grains should sublimate rapidly, and hence will not contribute to the opacity. On the other hand, \cite{2018MNRAS.481..703I} pointed out that small grains could be present if the optical depth is large across the whole of the transiting structures, transitioning with a rapid gradient to zero towards their edge, so there would be no noticeable optically-thin regime. As with \rev{these other WDs}, we see no colour dependence in the transits of WD~1054-226. For WD~1054-226, the longer orbital period and lower stellar $T_\mathrm{eff}$ mean that the grain temperature is much lower than for WD~1145+017: we calculate a blackbody grain temperature of 320~K at $P=25$~hr, while accounting for inefficient cooling of sub-micron grains \citep[][Eq 8]{2018MNRAS.474.4795X} gives an estimated grain temperature of 600~K. This is far too low for the grains to sublimate on relevant timescales: following \citep{2018MNRAS.474.4795X}, we find a sublimation lifetime of $10^{18}$~yr for 10~nm grains. This means that for WD~1054-226, sublimation cannot be invoked as a means to remove small grains. Instead, they must be present, but rendered invisible in the colour information presumably by high optical depth. \rev{We note that the nondetection of circumstellar gas absorption in WD~1054-226 by \cite{2022MNRAS.511.1647F}, in contrast to detections in WD~1145+017 \citep{2016ApJ...816L..22X,2017ApJ...839...42R}, would be consistent with a lack of sublimation in the former's disc.}

\section{Conclusions}

We have presented updated time-series photometry for the variable white dwarf WD~1054-226, including a new sector of \tess observations as well as multi-band ground-based photometry.

WD~1054-226 still displays significant variability with periods of 25.01~h and 23.1~min, which have remained stable now over 6 years, indicating a relatively long-lived origin. These periods appear both with a \rev{Lomb--Scargle} periodogram analysis, and when incorporating a \rev{Gaussian Process} to account for aperiodic variability. The 23.1~min signal remains at the 65:1 period commensurability with the 25.01~h signal, qualitatively consistent with the original interpretation of \cite{2022MNRAS.511.1647F}, which posits that they arise from the sculpting of a circumstellar disc by a nearby or embedded asteroid. However, in \tess Sector~36, the period ratio drifted from 65:1, which may indicate additional processes in the system, or cast doubt on this interpretation.

The ground-based observations included simultaneous multi-band observations, which showed no colour dependence of the transits, as is also the case for WD~1145+017 \rev{and several similar WDs}. However, for WD~1054-226, the transiting material is cool enough that rapid sublimation of small dust grains cannot be invoked. Instead, the disc of material may be optically thick at all passbands observed so far.

While a quantitative model for this system remains lacking, any such model must account for both the disc's high optical depth and the stability of the dominant frequencies.

\begin{acknowledgements}
This paper includes data collected with the \tess mission, obtained from the MAST data archive at the Space Telescope Science Institute (STScI). Funding for the \tess mission is provided by the NASA Explorer Program. STScI is operated by the Association of Universities for Research in Astronomy, Inc., under NASA contract NAS 5–26555. This work makes use of observations from the Las Cumbres Observatory global telescope network. This paper is based on observations made with the MuSCAT2 instrument, developed by ABC, at Telescopio Carlos Sánchez operated on the island of Tenerife by the IAC in the Spanish Observatorio del Teide. This paper includes data taken at the McDonald Observatory of the University of Texas at Austin. The data presented here were obtained in part with ALFOSC, which is provided by the Instituto de Astrofisica de Andalucia (IAA) under a joint agreement with the University of Copenhagen and NOT. The authors acknowledge support from the Swiss NCCR PlanetS and the Swiss National Science Foundation. This work has been carried out within the framework of the NCCR PlanetS supported by the Swiss National Science Foundation under grants 51NF40182901 and 51NF40205606.

JK acknowledges support of the Swiss National Science Foundation under grant number TMSGI2\_211697. AJM and JK acknowledge support from the Swedish Research Council (Starting Grant 2017-04945 and Project Grant 2022-04043). 
HP acknowledges support by the Spanish Ministry of Science and Innovation with the Ramon y Cajal fellowship number RYC2021-031798-I. VB acknowledges support from grant PID2022-137241NB-C41 funded by Agencia Estatal de Investigación of 
the  Ministerio de Ciencia, Innovación y Universidades (MICIU/AEI/10.13039/501100011033) and ERDF/EU. E. E-B. acknowledges financial support from the European Union and the State Agency of Investigation of the Spanish Ministry of Science and Innovation (MICINN) under the grant PRE2020-093107 of the Pre-Doc Program for the Training of Doctors (FPI-SO) through FSE funds. G.M. acknowledges financial support from the Severo Ochoa grant CEX2021-001131-S and from the Ramón y Cajal grant RYC2022-037854-I funded by MCIN/AEI/1144 10.13039/501100011033 and FSE+. 
Funding from the University of La Laguna and the Spanish Ministry of Universities is acknowledged. This work is partly financed by the Spanish Ministry of Economics and Competitiveness through grants PGC2018-098153-B-C31. We acknowledge financial support from the Agencia Estatal de Investigaci\'on of the Ministerio de Ciencia e Innovaci\'on MCIN/AEI/10.13039/501100011033 and the ERDF “A way of making Europe” through project PID2021-125627OB-C32. This work is partly supported by JSPS KAKENHI Grant Numbers JP21K13955, JP24H00017, JP24H00248, JP24K00689, JP24K17082, and JP24K17083, JSPS Bilateral Program Number JPJSBP120249910, JST SPRING Grant Number JPMJSP2108, and JSPS Grant-in-Aid for JSPS Fellows Grant Numbers JP24KJ0241, JP25KJ0091, and JP25KJ1040.
\end{acknowledgements}

\bibliographystyle{aa} 
\bibliography{example} 

\begin{appendix}
\onecolumn

\section{Observation Summary}
\label{sec:observing_log}

\begin{table}[ht!]
    \centering
    \caption{Observation log.}
    \begin{tabular*}{\textwidth}{@{\extracolsep{\fill}} cccccc}
    \toprule
    \toprule
    Start Date & $N_\mathrm{exp}$ & Exp. time & \rev{Duration} & Filter & Instrument \\\relax
    [UTC] & & [s] & \rev{[h]} & & \\
    \midrule
    2022-12-14 & 89 & 60 & 1.48 & $i'$ & Sinistro\\
    2022-12-29 & 166 & 60 & 2.77 & $i'$ & Sinistro\\
    2023-01-24 & 449, 869, 449, 449 & 30, 15, 30, 30 & 3.74 & $g',r',i,z_s$ & MuSCAT2\\
    2023-01-26 & 111, 832, 111, 111 & 120, 15, 120, 120 & 3.70 & $g',r',i,z_s$ & MuSCAT2\\
    2023-01-30 & 208 & 60 & 3.47 & $i'$ & Sinistro\\
    2023-01-31 & 144 & 60 & 2.40 & $i'$ & Sinistro\\
    2023-03-18 & 28, 30 & 120, 120 & 1.00 & $g',i'$ & ALFOSC\\
    2024-01-05 & 333, 332  & 10, 10 & 0.93 & $g',i'$ & ProEM\\
    2024-01-06 & 480, 477 & 10, 10 & 1.33 & $g',i'$ & ProEM\\
    2024-01-06 & 196 & 60 & 3.27 & $i'$ & Sinistro\\
    2024-01-06 & 201 & 60 & 3.35 & $i'$ & Sinistro\\
    2024-01-07 & 198 & 60 & 3.30 & $i'$ & Sinistro\\
    2024-01-09 & 188 & 60 & 3.13 & $i'$ & Sinistro\\
    2024-01-10 & 243 & 60 & 4.05 & $i'$ & Sinistro\\
    2024-01-11 & 244 & 60 & 4.07 & $i'$ & Sinistro\\
    2024-01-12 & 247 & 60 & 4.12 & $i'$ & Sinistro\\
    2024-01-12 & 236 & 60 & 3.93 & $i'$ & Sinistro\\
    2024-01-13 & 254 & 60 & 4.23 & $i'$ & Sinistro\\
    2024-01-23 & 291 & 60 & 4.85 & $i'$ & Sinistro\\
    2024-01-23 & 293 & 60 & 4.88 & $i'$ & Sinistro\\  
    2024-01-24 & 299 & 60 & 4.98 & $i'$ & Sinistro\\
    2024-01-24 & 183 & 60 & 3.05 & $i'$ & Sinistro\\  
    2024-01-28 & 209 & 60 & 3.48 & $i'$ & Sinistro\\
    2024-01-31 & 128 & 60 & 2.13 & $i'$ & Sinistro\\
    2024-03-07 & 449 & 60 & 7.48 & $i'$ & Sinistro\\
    2024-03-08 & 445 & 60 & 7.42 & $i'$ & Sinistro\\
    2024-03-08 & 432 & 60 & 7.20 & $i'$ & Sinistro\\
    2024-03-08 & 208, 790, 208, 201 & 60, 15, 60, 60 & 3.47 & $g',r',i,z_s$ & MuSCAT2\\
    2024-03-09 & 438 & 60 & 7.30 & $i'$ & Sinistro\\
    2024-03-09 & 231, 879, 231, 231 & 60, 15, 60, 60 & 3.85 & $g',r',i,z_s$ & MuSCAT2\\
    2024-03-09 & 432 & 60 & 7.20 & $i'$ & Sinistro\\
    2024-03-10 & 201, 765, 201, 202 & 60, 15, 60, 60 & 3.35 & $g',r',i,z_s$ & MuSCAT2\\
    2024-03-11 & 175, 662, 175, 175 & 60, 15, 60, 60 & 2.92 & $g',r',i,z_s$ & MuSCAT2\\
    2024-03-12 & 121, 459, 121, 121 & 60, 15, 60, 60 & 2.02 & $g',r',i,z_s$ & MuSCAT2\\
    2025-06-20 & 176 & 60 & 2.93 & $i'$ & Sinistro\\
    2025-06-21 & 175 & 60 & 2.92 & $i'$ & Sinistro\\
    2025-06-23 & 167 & 60 & 2.78 & $i'$ & Sinistro\\
    2025-06-24 & 158 & 60 & 2.63 & $i'$ & Sinistro\\
    2025-06-27 & 158 & 60 & 2.63 & $i'$ & Sinistro\\
    \bottomrule
    \end{tabular*}
    \label{tab:observation_summary}
\end{table}

\clearpage

\section{MuSCAT2 Multicolour Photometry}
\label{sec:m2_appendix}
\begin{figure}[ht!]
    \centering
    \includegraphics[width=0.95\linewidth]{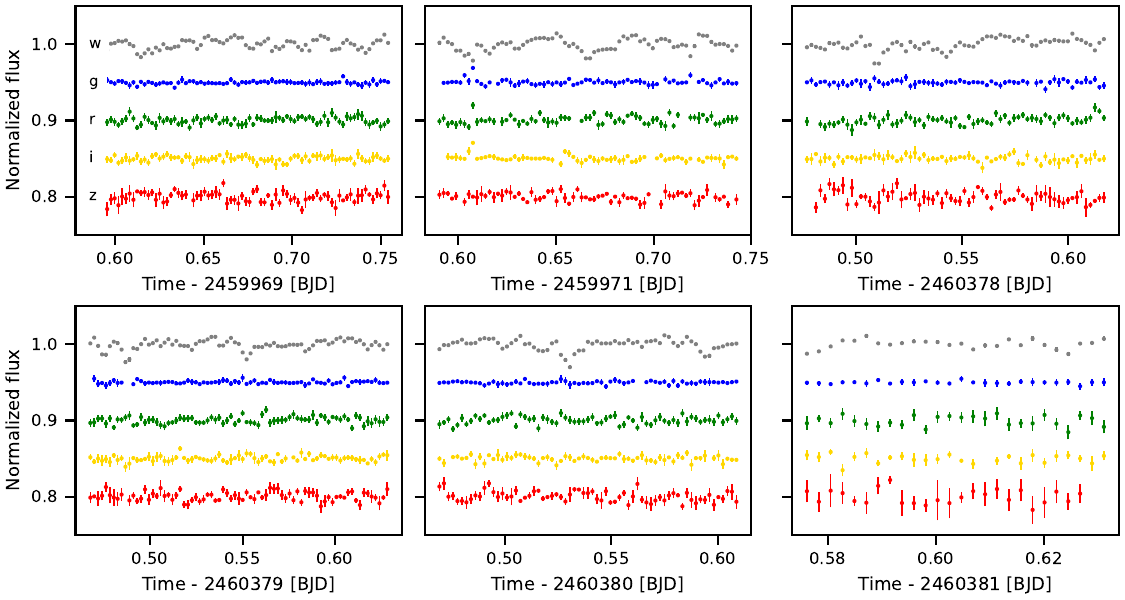}
    \caption{White light curves and the colour residuals against g, r, i, and z, light curves for six MuSCAT2 multicolour light curves. One of the observed light curves is excluded due to poor photometric quality.}
    \label{fig:M2_white}
\end{figure}

\begin{figure}[ht!]
    \centering
    \includegraphics[width=0.95\linewidth]{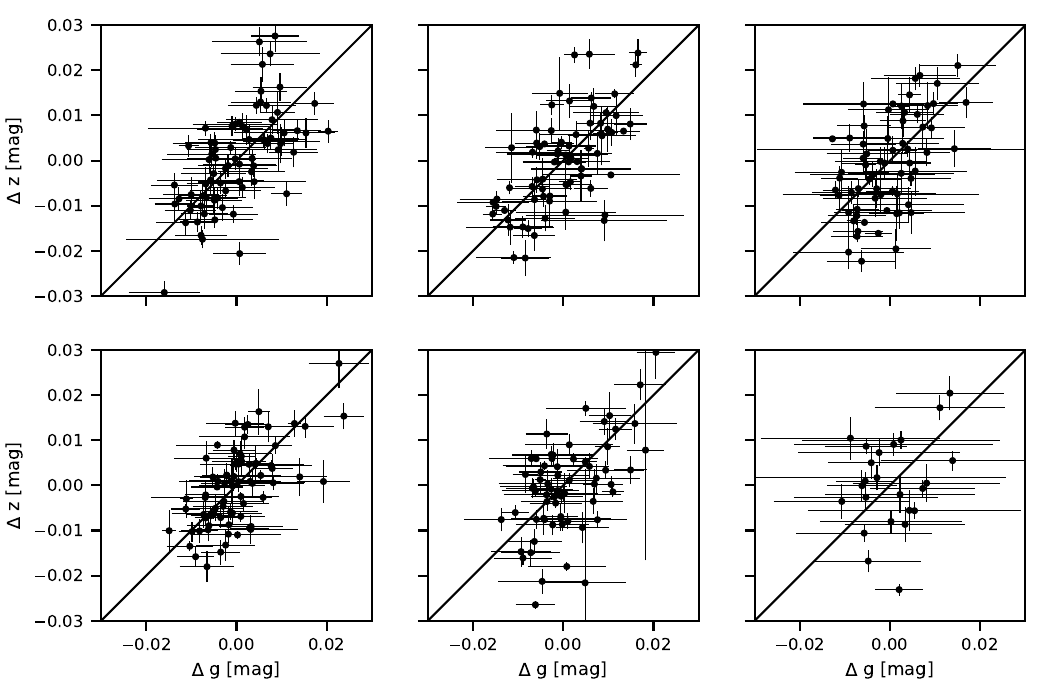}
    \caption{\rev{MuSCAT2 z- versus g-band light curves (in $\Delta$~mag). We do not detect any systematic colour signatures within uncertainties.}}
    \label{fig:m2_zg}
\end{figure}
    
\end{appendix}

\end{document}